\documentclass[12pt,a4paper]{article}
\pdfoutput=1
\usepackage{jheppub}
\usepackage[utf8]{inputenc}
\usepackage[T1]{fontenc}
\usepackage{comment}

\usepackage{amsmath,physics,float}  
\usepackage{amssymb}  
\usepackage{latexsym}
\usepackage{graphicx,xcolor}
\usepackage{cleveref}

\usepackage{amsfonts}
\usepackage{braket}
\usepackage{mathrsfs}

\usepackage{subfig} 
\usepackage{empheq}

\usepackage[shortlabels]{enumitem}

\usepackage{pgfplots}
\usepackage{natbib} 
\usepackage{bbm}

\usepackage{bigints}

\newcommand{\beq}{\begin{equation}}
\newcommand{\eeq}{\end{equation}}

\usepackage{tikz, pgf}
\usetikzlibrary{decorations.pathmorphing}
\usetikzlibrary{arrows.meta}
\tikzset{%
  >={Latex[width=2mm,length=2mm]},
            base/.style = {rectangle, rounded corners, draw=black,
                           minimum width=4cm, minimum height=1cm,
                           text centered, font=\sffamily},
}
\usetikzlibrary{calc,bending,decorations.markings}
\begin{document}

\title{\boldmath On the positivity of Coon amplitude in $D=4$}

\author[]{Joydeep Chakravarty,}
\emailAdd{joydeep.chakravarty@icts.res.in}

\author[]{Pronobesh Maity, and}
\emailAdd{pronobesh.maity@icts.res.in}

\author[]{Amiya Mishra}
\emailAdd{amiya.mishra@icts.res.in}

\affiliation[]{International Centre for Theoretical Sciences-TIFR,\\
Shivakote, Hesaraghatta Hobli,
Bangalore 560089, India.}
\vspace{4pt}

\begin{abstract}
{The Coon amplitude is the unique solution to duality constraints with logarithmic Regge trajectories. A striking feature of this solution is that it interpolates between the Veneziano amplitude and a scalar particle amplitude. However, an analytic proof of unitarity of the amplitude is not yet known. In this short note, we explicitly compute the partial wave coefficients on the leading Regge trajectory in $D=4$. We find that these coefficients always remain positive, even though their magnitude decreases with spin. Since the coefficients on the subleading trajectories are observed to be larger than those on the leading ones, our result indicates the positivity of the full Coon amplitude in $D=4$.}
\end{abstract} 

\maketitle

\section{Introduction}
The Coon amplitude is a $2 \to 2$ scattering amplitude describing an infinite number of higher spin exchanges consistent with the duality constraints \cite{Coon:1969yw, Baker:1970vxk, Coon:1972qz}. In particular, this mysterious amplitude is the unique solution to duality constraints having an infinity of exchanges with logarithmic trajectories. Recently, it has been brought forward as an interesting case study on its own \cite{Ridkokasha:2020epy, Figueroa:2022onw, Geiser:2022icl, Maldacena:2022ckr, Caron-Huot:2016icg} while also serving as a useful reference to the study of accumulation point amplitudes with similar infinite towers of states \cite{Caron-Huot:2016icg, Bern:2021ppb, Huang:2022mdb, Arkani-Hamed:2020blm, Chiang:2022jep}. 

An intriguing feature of the amplitude is that it interpolates between the stringy and the worldline descriptions, i.e., between the Veneziano amplitude and a scalar amplitude. The expression for the amplitude is as follows:
\begin{equation}\label{Coon}
	A_q(s,t)=(q-1)\,q^{\frac{\log\sigma \log\tau}{(\log q)^2}} \prod_{n=0}^{\infty} \frac{(\sigma\tau-q^n)(1-q^{n+1})}{(\sigma-q^n)(\tau-q^n)}
	\end{equation}
where $q$ is a parameter ranging from $q\in(0,1)$. As it may be apparent, the Coon amplitude can be thought of as a $q$-deformation of the Veneziano amplitude \cite{Chaichian:1992hr, Jenkovszky:1994qg}. Here the parameters $\sigma$ and $\tau$ can be expressed in terms of the Mandelstam variables $s$ and $t$, respectively:
\beq\label{sigma_tau}
\sigma=1+(s-m^2)(q-1),\quad \tau=1+(t-m^2)(q-1).
\eeq
As we approach the limiting values of $q$, the amplitude reduces to scalar particle and Veneziano amplitudes, respectively (see Appendix \ref{A} for further details regarding the limiting cases):
	\begin{equation}\label{limits}
	\begin{split}
		&\lim_{q\to 0} A_q(s,t)=\frac{1}{s-m^2}+\frac{1}{t-m^2}-1
		\\
		& \lim_{q\to 1} A_q(s,t) = - \frac{\Gamma(-s+m^2)\Gamma(-t+m^2)}{\Gamma(-s-t+2m^2)}
			\end{split}
	\end{equation}
The Regge trajectories corresponding to the amplitude for a generic value of $q$ are logarithmic. The amplitude (\ref{Coon}) has real $s$ poles at the locations:
\begin{equation} \label{coonpoles}
\sigma =q^n \;\leftrightarrow \; s=m^2+\frac{1-q^n}{1-q}.
\end{equation} 
Equation \eqref{coonpoles} indicates the existence of an accumulation point, i.e., a finite energy scale which the Regge poles asymptote to upon taking the limit $n\to \infty$ of \eqref{coonpoles}.
\begin{equation} \label{accupt}
s_*=m^2+\frac{1}{1-q}
\end{equation}
Apart from certain aspects of asymptotic behaviour, not much about this amplitude is well understood. In particular, details regarding the underlying physical interpretation of the amplitude are lacking, although a proposal was recently put forward regarding the same \cite{Maldacena:2022ckr}. Additionally, from previous works \cite{PhysRevD.10.3780, Fairlie:1994ad}, it is not clear whether the amplitude \eqref{Coon} obeys unitarity constraints, though a recent numerical analysis indicates that it is so \cite{Figueroa:2022onw}.\footnote{Historically, the conclusions of \cite{PhysRevD.10.3780} were instrumental in abandoning the further study of the Coon amplitude, since their numerical work indicated the presence of ghosts, i.e., negative norm states. Our analytical results on positivity indicate the absence of such states at tree-level in $D=4$, confirming the analysis of \cite{Figueroa:2022onw}.}

The main aim of our work is to develop an understanding of the unitarity of the amplitude \eqref{Coon}. One of the important implications of unitarity is the positivity of the coefficients of the partial wave expansion. In our work, we compute the partial wave coefficients corresponding to the leading Regge trajectory of the Coon amplitude in $D=4$ (see equation \eqref{2.26}).  We find that the coefficients are always positive, and their magnitude gradually decreases with spin. In addition, we also numerically find that the partial wave coefficients on the subleading trajectories are larger than the coefficients corresponding to the leading Regge trajectory above sufficiently large spin values (see Appendix \ref{AC}). This indicates the consistency of the amplitude with unitarity in $D=4$. In Appendix \ref{A3}, we determine the general form of the partial wave coefficient corresponding to the subleading trajectories (see equation \eqref{B.3}). 

\section{Positivity of the Coon amplitude}
Prior to the computation, we simplify the notational baggage to make our expressions visually less demanding. 
\subsection{Notation}
Firstly let us define the $q$-Pochhammer symbol as follows
\begin{equation} \label{2.1}
	(a;q)_{N}=\prod_{n=0}^{N-1}(1-aq^{n}).
\end{equation} 
Next we define the variables $\alpha_s$ and $\alpha_t$ which are functions of the Mandelstam variables $s$ and $t$ as given below 
\beq\label{2.2}
\alpha_s=\frac{\log\sigma}{\log q},\quad \alpha_s=\frac{\log\tau}{\log q}.
\eeq
We will also introduce $[n]_q$ as a q-deformed integer with the following expression:
\beq\label{2.3}
[n]_q=\frac{1-q^n}{1-q}.
\eeq
Note that as $q\to 1$, the q-deformed integer reduces to ordinary numbers $n$. Using equations \eqref{2.1}, \eqref{2.2} and \eqref{2.3}, we can conveniently express the Coon amplitude (\ref{Coon}) in a compact form
\begin{equation}\label{2.4}
A_{q}(s,t) =(q-1)\,q^{\alpha_s\,\alpha_t} \frac{(q^{-\alpha_s-\alpha_t};q)_{\infty}(q;q)_{\infty}}{(q^{-\alpha_s};q)_{\infty}(q^{-\alpha_t};q)_{\infty}}
\end{equation}
The real $s$ poles given in \eqref{coonpoles} now take the following form:
\begin{equation} \label{2.5}
s=m_n^2\equiv m^2+[n]_q \;\leftrightarrow \; \sigma =q^n.
\end{equation}

\subsection{Partial wave decomposition}
\begin{figure}[!htb]
\centering
\includegraphics[width=.5\textwidth]{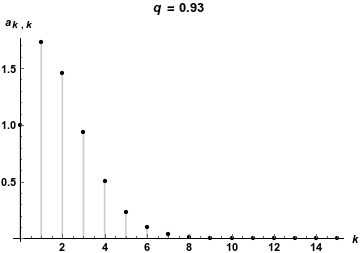}\hfill
\caption{Decreasing magnitude of leading Regge coefficients with spin for $q=0.93$ and $ m^2=-1$.}
\label{fig1}
 \end{figure}

\begin{figure}[!htb]
     \centering
    \includegraphics[width=.5\textwidth]{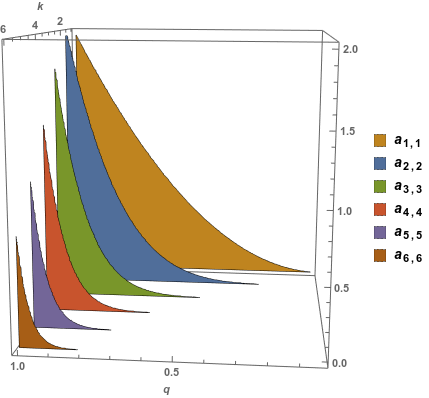}\hfill
     \caption{Plot displaying behaviour of leading Regge coefficients with $q$ and spin (for $m^2=-1$).}
     \label{fig:my_label}
 \end{figure}

The residue of (\ref{Coon}) at the pole $s=m_k^2$ from \eqref{2.5} is 
\begin{equation} \label{2.6}
	\text{Res}_{s=m_k^2}A_q(s,t)=\tau^k \left[\prod_{\substack{n=0 \\ n\neq k}}^{\infty} \frac{(q^k \tau-q^n)(1-q^{n+1})}{(q^k-q^n)(\tau-q^n)}\right]\times \frac{q^k(\tau-1)}{\tau-q^k}(1-q^{k+1}).
\end{equation}
Alternatively, using the $q$-Pochhammer symbol, the expression for the residue takes the following form: 
\begin{equation}\label{Residue}
	\begin{split}
	R_{k}(\tau,q)\equiv	\text{Res}_{s=m_k^2}A_q(s,t)&=\tau^k \, \frac{(1/(q^k\tau);q)_{\infty}(q;q)_{\infty}}{(1/\tau;q)_{\infty}}\times \lim_{\sigma \to q^k}\frac{(\sigma-q^k)}{(1/\sigma;q)_{\infty}}\\
		&=  \frac{(q;q)_{\infty}q^k }{\prod_{\substack{n=0 \\ n\neq k}}^{\infty}(1-q^{n-k})}\times\left[\tau^k\frac{\left(\frac{1}{\tau}\frac{1}{q^k};q\right)_{\infty}}{\left(\frac{1}{\tau};q\right)_{\infty}}\right].
	\end{split}
\end{equation}
We can decompose the product in the denominator in the following useful fashion:
\begin{equation}  \label{2.8}
\begin{split}
\prod_{\substack{n=0 \\ n\neq k}}^{\infty}(1-q^{n-k})=\prod_{n=0}^{k-1}(1-q^{n-k}) \prod_{n=0}^{\infty}(1-q^{n+1})=(q;q)_{\infty}\prod_{n=0}^{k-1}(1-q^{n-k}).
\end{split}
\end{equation}  
Using equation \eqref{2.8}, the residue \eqref{Residue} simplifies to give the expression:
\begin{equation}  \label{2.9}
R_{k}(\tau,q)=  \frac{q^k }{\prod_{n=0}^{k-1}(1-q^{n-k})}\times\left[\tau^k\frac{\left(\frac{1}{\tau}\frac{1}{q^k};q\right)_{\infty}}{\left(\frac{1}{\tau};q\right)_{\infty}}\right]
\end{equation}
To get the partial wave coefficients, we need to expand the residue obtained in \eqref{2.9} in terms of a partial wave decomposition. We decompose the residue in the following form in $D=4$:
\begin{equation}
	R_{k}(\tau,q)=\sum_{l=0}^{k} \, a_{k,l}^q\, P_l\left(1+\frac{2t}{[k]_q-3m^2} \right)
\end{equation}
Note that the Legendre polynomials $P_l(x)$ entering the partial wave expansion satisfy the following orthogonality relation:
\begin{equation} \label{ortho}
	\int_{-1}^{+1}\, dx\,P_l(x)\,P_{l'}(x)=\frac{2}{2l+1}\,\delta_{l,l'}
\end{equation}
We can use the orthogonality relation given in \eqref{ortho} to read off the partial wave amplitudes as
\begin{equation} \label{2.12}
	a_{k,l}^q=\frac{2l+1}{[k]_q-3m^2} \int_{t=3m^2-[k]_q}^{t=0} dt\, R_{k}(\tau,q)\, P_l\left(1+\frac{2t}{[k]_q-3m^2} \right)
\end{equation}
We have plotted $a^q_{k,k}$ for first few values of $k$ in figures \ref{fig1} and \ref{fig:my_label}. They decrease with spin $k$ but remain positive. 

\subsection{Calculation of residues}
Our method for analytically determining the residues obtained in \eqref{2.12} involves utilizing the generating function of the Legendre polynomials $P_l(x)$:
\begin{equation} \label{2.13}
	\sum_{l=0}^{\infty} P_l(x)\,h^{-l-1}=(1-2xh+h^2)^{-1/2}.
\end{equation}
Using the generating function of the polynomials in \eqref{2.13}, we can construct a set of generating functions for the partial wave coefficients $a_{n,j}$ by dividing with $(2j+1)h^{j+1}$ on both sides and then taking the sum over all spins $j$:
\begin{equation} \label{2.14}
	G_{k}(h)=\sum_{j=0}^{\infty} \frac{1}{2j+1} \frac{a^q_{k,j}}{h^{j+1}}=\frac{1}{[k]_q-3m^2}\int_{t=3m^2-[k]_q}^{t=0} dt\, \frac{R_k(\tau,q)}{\left[(h-1)^2-\frac{4ht}{[k]_q-3m^2}\right]^{1/2}}
\end{equation}
In order to compute \eqref{2.14}, we need to expand the original expression for the residue given in (\ref{Residue}) in terms of the variable $\tau$. This operation necessitates the usage of the following identity
 \begin{equation}\label{A1}
 \frac{(az;q)_{\infty}}{(z;q)_{\infty}} = \sum_{n=0}^{\infty} \frac{(a;q)_n}{(q;q)_n}\, z^n ,
 \end{equation}
where we have used the notation introduced in \eqref{2.1}. Then, the identity \eqref{A1} helps us in evaluating a part of the residue from (\ref{Residue}) as given below:
\begin{equation} \label{2.16}
		\frac{\left(\frac{1}{\tau}\frac{1}{q^k};q\right)_{\infty}}{\left(\frac{1}{\tau};q\right)_{\infty}}=\sum_{n=0}^{\infty}	\frac{\left(\frac{1}{q^k};q\right)_{n}}{\left(q;q\right)_{n}} \,\frac{1}{\tau^n}=\sum_{n=0}^{k}	\frac{\left(\frac{1}{q^k};q\right)_{n}}{\left(q;q\right)_{n}} \,\frac{1}{\tau^n}
\end{equation}
where in the second equality, we have restricted the upper limit of the product from $\infty$ to $k$ using the fact that the numerator vanishes for $n>k$, i.e.
\begin{equation}
\left(\frac{1}{q^k};q\right)_{n} =0\;\, \text{for} \; n>k 
\end{equation}
With the simplification introduced in \eqref{2.16} the residue given in \eqref{Residue} takes the following form:
\begin{equation} \label{2.18}
 R_k(\tau,q)= \frac{q^k }{\prod_{n=0}^{k-1}(1-q^{n-k})} \sum_{n=0}^{k}	\frac{\left(\frac{1}{q^k};q\right)_{n}}{\left(q;q\right)_{n}} \, \tau^{k-n}
\end{equation}
\subsubsection{Bound on spin}
\label{subsub}
As clear from the expression \eqref{2.18}, the residue at a given pole $s=m_k^2$ is a polynomial in $t$ of degree $k$, corresponding to particle exchanges of spin $j=0,\dots,k$. Note that the poles of our amplitude $s=m^2+[k]_q$ start from $k=0$, as opposed to  \cite{Maity:2021obe} where the poles ($s=k$) start from $k=-1$. Thus we need to set $m^2 =-1$ and shift $k\to (k+1)$ in the expressions of $a^q_{k,j}$ to compare our results to \cite{Maity:2021obe}. In our convention, the leading Regge trajectory is associated with coefficients of the form $a^q_{k,k}$. 

\subsection{Reading off the coefficients for leading Regge trajectories}
Using the simplified residues from \eqref{2.18}, the expression for the generating function $G_k(h)$ given in \eqref{2.14} takes the following form.
\begin{equation} \label{2.19}
	\begin{split}
G_k(h)= \frac{1}{[k]_q-3m^2} \frac{q^k }{\prod_{n=0}^{k-1}(1-q^{n-k})}\sum_{n=0}^{k}	\frac{\left(\frac{1}{q^k};q\right)_{n}}{\left(q;q\right)_{n}}  \int_{t=3m^2-[k]_q}^{t=0} dt\, \frac{\tau^{k-n}}{\left[(h-1)^2-\frac{4ht}{[k]_q-3m^2}\right]^{1/2}}
\end{split}
\end{equation}
As can be seen from \eqref{2.14}, in order to obtain the expression from  $a^q_{k,j}$, we have to read off the coefficient of $h^{-j-1}$ from the expression. To extract this coefficient, we can define a \textit{pseudo-generating function} $g_{k,r}(h)$ in the subsequent fashion:
\begin{equation} \label{2.20}
g_{k,r}(h)=\int_{t=3m^2-[k]_q}^{t=0} dt\, \frac{\tau^{r}}{\left[(h-1)^2-\frac{4ht}{[k]_q-3m^2}\right]^{1/2}}.
\end{equation}
Using the definition introduced in \eqref{2.20}, the expression for the original generating function $G_k(h)$ from \eqref{2.19} has the expression
\begin{equation}\label{q}
	G_k(h)=\frac{1}{[k]_q-3m^2} \frac{q^k }{\prod_{n=0}^{k-1}(1-q^{n-k})}\sum_{r=0}^{k}	\frac{\left(\frac{1}{q^k};q\right)_{k-r}}{\left(q;q\right)_{k-r}} \, g_{k,r}(h)
\end{equation}
The pseudo generating function introduced in \eqref{2.20} can be explicitly evaluated and can be expressed as a sum of two ${}_{2}F_1$ hypergeometric functions:
\begin{equation}\label{g[n,k]1} 
\begin{split}
g_{k,r}(h)&=\frac{(q-1)^{r}}{2^{2r+1}}\, \frac{([k_q]-3m^2)^{r+1}}{h^{r+1}}  \left[(h-1)^2+\frac{4hs_*}{3m^2-[k]_q}\right]^r \times \\&\Big[ (h+1) {}_2F_{1}\left( \frac{1}{2},-r;\frac{3}{2};\frac{(h+1)^2}{(h-1)^2+\frac{4hs_*}{3m^2-[k]_q}} \right)-(h-1){}_2F_{1}\left( \frac{1}{2},-r;\frac{3}{2};\frac{(h-1)^2}{(h-1)^2+\frac{4hs_*}{3m^2-[k]_q}}\right)\Big].
\end{split}
\end{equation}
where $s_*$ is the accumulation point calculated in \eqref{accupt}.

\subsubsection{Reading off the leading Regge trajectory}
The pseudo generating function $g_{k,r}(h)$ comprises of $(r+1)$ terms accompanying powers of $h$, i.e. terms of the form $h^{-1}, h^{-2},\cdots, h^{-r-1}$. We can easily extract the coefficient of $h^{-r-1}$ from the expansion \eqref{g[n,k]1}, which gives us the following expression.
\begin{equation}\label{d}
 d_{k,r}= \frac{\sqrt{\pi}}{2^{2r+1}}\,\frac{r!}{\Gamma(r+3/2)}(q-1)^{r} ([k_q]-3m^2)^{r+1}
\end{equation}
In order to read off the partial wave coefficients, we will use a similar argument to the one used while extracting the leading Regge coefficients for the Veneziano amplitude \cite{Maity:2021obe}. Firstly, we note that the $h$-expansion of $G_k(h)$ can be represented in the following form using equations \eqref{2.14}, \eqref{q} and the expansion of $g_{k,r}(h)$ in $h$:
\begin{equation}\label{rewrite}
\underbrace{\sum_{j=0}^{k}[...]\,h^{-j-1}}_{G_k(h)}=\sum_{r=0}^{k}[...]\,\underbrace{\sum_{p=0}^{r}[...]\,h^{-p-1}}_{g_{k,r}(h)}.
\end{equation}
In general, it is difficult to extract the coefficients $a^q_{k,l}$ from \eqref{rewrite}. However, we can determine the expression for the leading Regge trajectory, i.e., $a^q_{k,k}$. In order to determine $a^q_{k,k}$ from the coefficient of $h^{-k-1}$ in the RHS of \eqref{rewrite}, we need to consider the term with $h^{-k-1}$ in $g_{k,k}(h)$, which is a unique term. Using this term, while equating both sides of \eqref{rewrite}, we obtain the following expression for $a^q_{k,k}$:
\begin{equation} \label{2.25}
\begin{split}
\frac{1}{2k+1}\,a^q_{k,k}=\frac{1}{[k]_q-3m^2} \frac{q^k }{\prod_{n=0}^{k-1}(1-q^{n-k})}	\frac{\left(\frac{1}{q^k};q\right)_{0}}{\left(q;q\right)_{0}}\times d_{k,k} 
\end{split}
\end{equation}
Substituting the result of $d_{k,k}$ from (\ref{d}) into \eqref{2.25}, we obtain the our result for the partial wave coefficients $a^q_{k,k}$
\begin{equation} \label{2.26}
	a^q_{k,k}=\frac{\sqrt{\pi}}{2^{2k}} \frac{k!}{\Gamma(k+1/2)}\frac{([k]_q-3m^2)^{k}q^k (1-q)^{k}}{\prod_{n=0}^{k-1}(q^{n-k}-1)}.
\end{equation}
Equation \eqref{2.26} reproduces the plot in figure \ref{fig:my_label}, and is manifestly positive for $m^2<\frac{1}{3}\text{min}\,([k]_q)=\frac{1}{3}$ \footnote{Note that for $k=0$, $a^q_{0,0}=1>0$.} and $0<q<1$. This upper bound on $m^2$ is consistent with the bound obtained in \cite{Figueroa:2022onw}. 

As another check, we take the $q\to 1$ limit of equation \eqref{2.26}, where we expect to recover the coefficients for the Veneziano amplitude. In the limit $q\to 1$, we have the following reduction:
\beq \label{2.27}
\prod_{n=0}^{k-1}(q^{n-k}-1)\to (1-q)^k k!.
\eeq
Taking the limit $q\to 1$ in \eqref{2.26} and substituting \eqref{2.27} in the same, we obtain
\begin{equation}
	\lim_{q\to 1} a^q_{k,k} = \frac{\sqrt{\pi}}{2^{2k}} \frac{(k-3m^2)^{k}}{\Gamma(k+1/2)}
\end{equation}
 which precisely matches (accounting for the shifts in \ref{subsub}) with the result for Veneziano amplitude in \cite{Maity:2021obe, Arkani-Hamed:2022gsa}.

\section{Conclusion}
In our work, we have calculated the partial wave coefficients that govern the leading Regge behaviour in $D=4$. We have shown that the associated coefficients are positive. Since the coefficients governing the subsequent subleading Regge trajectories have increasingly larger values, our work indicates the positivity of the Coon amplitude for all partial wave coefficients in $D=4$. An interesting direction would be to investigate the same for general $D$, possibly along the lines of \cite{Arkani-Hamed:2022gsa}. 

In general, negative partial wave coefficients imply that exchanges involving negative norm states take place. Our present work indicates the absence of such states at the tree level. However, it may still be possible that the underlying physical description suffers from problems resulting from negative norm states when loops are taken into account, and hence the physical description is inconsistent in the first place. In order to check such issues, one needs to understand the physics behind the amplitude, which may be similar to the prescription suggested by \cite{Maldacena:2022ckr}. We aim to understand aspects of the underlying description in a forthcoming work.

The accumulation point in \eqref{accupt} leads to the existence of a branch cut, and another interesting question is to understand the physical structure behind the same. 

 \begin{acknowledgments}
We thank members of the string group at ICTS Bangalore for various discussions, and valuable communications regarding our work, especially Rajesh Gopakumar and  Ashoke Sen. We also thank Diksha Jain, R Loganayagam, Arnab Rudra, Akhil Sivakumar, and Spenta Wadia for related discussions. We acknowledge the support of the Department of Atomic Energy, Government of India, under project number RTI4001. The authors also acknowledge gratitude to the people of India for their steady and generous support to research in basic sciences.

\end{acknowledgments}
\appendix

\section{Expression for the general Regge coefficient $a^q_{k,j}$}
\label{A3}
Using the following expansion of the hypergeometric function 
\begin{equation}
{}_2F_{1} \left(\frac{1}{2},-r;\frac{3}{2};z\right)=\sum_{p=0}^{r} \frac{(-r)_p}{(2p+1)\, p!}\, z^p ,
\end{equation}
we can expand the expression for the pseudo-generating function $g_{k,r}(h)$ given in \eqref{g[n,k]1} as follows:
\begin{equation}\label{g[n,k]2}
\begin{split}
   g_{k,r}(h)=\frac{(q-1)^{r}}{2^{2r+1}}\, \frac{([k]_q-3m^2)^{r+1}}{h^{r+1}}&\sum_{p=0}^{r} \frac{(-r)_p}{(2p+1)p!} \left[(h-1)^2+\frac{4hs_*}{3m^2-[k]_q}\right]^{r-p} \\ &\times \sum_{d=0}^{2p+1} \binom{2p+1}{d} \,h^d (1+(-1)^d).  \\
\end{split}
\end{equation}
Now we can make binomial expansion of $\left[(h-1)^2+\frac{4hs_*}{3m^2-[k]_q}\right]^{r-p}$ in the above expression and compute the coefficient of $h^{-j-1}$ from r.h.s of \eqref{q} to find the generic coefficient on any Regge trajectory as
\begin{equation}\label{B.3}
    \begin{split}
        a^q_{k,j}=\frac{(2j+1)\,q^k}{\prod_{n=0}^{k-1}(1-q^{n-k})}\sum_{r=0}^{k}&\frac{1}{2^{2r+1}} \frac{\left(\frac{1}{q^k};q\right)_{k-r}}{\left(q;q\right)_{k-r}} (q-1)^r([k]_q-3m^2)^{r}\\
        &\times \sum_{p=0}^{r}\sum_{d=0}^{2p+1}\sum_{i=0}^{r-p} \binom{r-p}{i}\binom{2p+1}{d}\binom{2m}{p+i-j-d}\, \frac{(-r)_p}{(2p+1)\,p!}\\
        &\times \left(\frac{4s_*}{3m^2-[k]_q}\right)^{r-p-i}(-1)^{p+i-j-d}\,[1+(-1)^d]
    \end{split}
\end{equation}
Algorithmically \eqref{B.3} is much simpler than the original expression \eqref{2.12} as an integral.

\section{Leading versus subleading coefficients}
\label{AC}

In this appendix, we compare the partial wave coefficients on the leading Regge trajectory ($a^q_{k,k}$) versus the coefficients on the subleading Regge trajectory ($a^q_{k,j}, \, j<k$). 

\begin{figure}[!htb] 

\centering
\includegraphics[width=.4\textwidth]{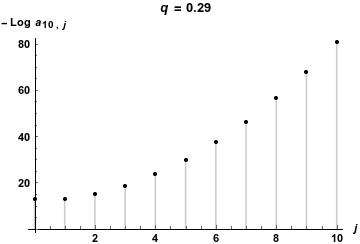}\hfill
\includegraphics[width=.4\textwidth]{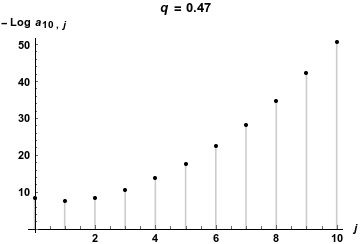}\hfill

\vspace{4mm}

\includegraphics[width=.4\textwidth]{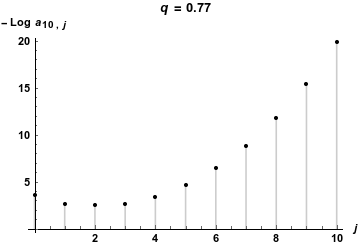}\hfill
\includegraphics[width=.4\textwidth]{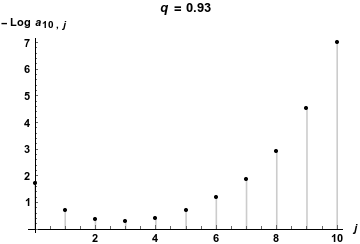}\hfill
\caption{Coefficients $a^q_{10,j}$ for $j=0,\dots,10$ at different values of $q$ for $m^2 =-1$.} \label{fig33}
\end{figure}

In figure \ref{fig33}, we perform the comparison between leading and subleading coefficients by fixing $k=10$, and plotting the negative logarithm of all coefficients $j$ with $j \leq 10$, for $m^2 =-1$. The increase in the negative logarithm of $a^q_{10,j}$ with $j$ implies that the magnitude of the coefficients $a^q_{10,j}$ decreases with $j$.

\begin{figure}[!htb] 

\centering
\includegraphics[width=.5\textwidth]{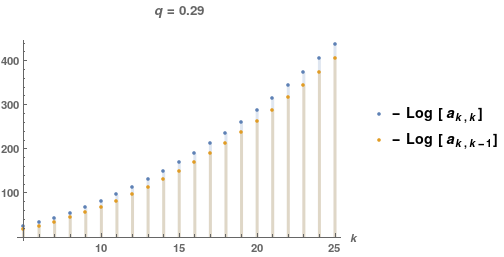}\hfill
\includegraphics[width=.5\textwidth]{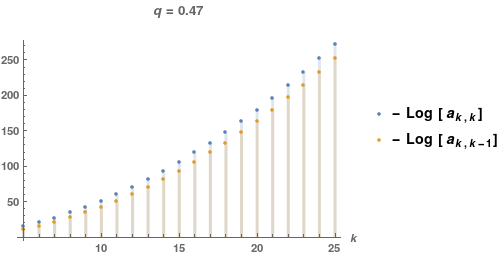}\hfill

\vspace{4mm}

\includegraphics[width=.5\textwidth]{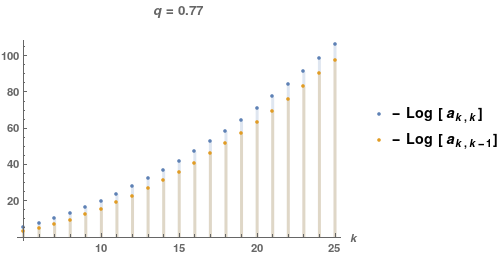}\hfill
\includegraphics[width=.5\textwidth]{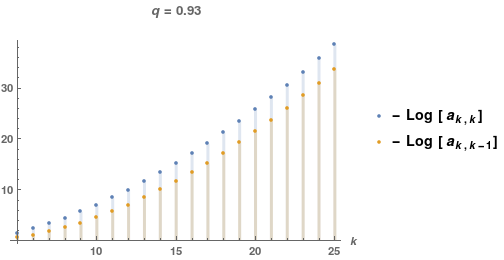}\hfill
\caption{Coefficients on the leading ($a^q_{k,k}$) and subleading ($a^q_{k,k-1}$) trajectories at different values of $q$ for $m^2 =-1$. }
\label{fig44}
 \end{figure}

 We also plot the coefficients corresponding to the leading ($a^q_{k,k}$) and the subleading ($a^q_{k,k-1}$) trajectories at different values of $q$ in figure \ref{fig44}. Here again, the increase in the negative logarithm of $a^q_{k,j}$ with $j$ implies that the magnitude of the coefficients $a^q_{k,j}$ decreases with $j$.
 
  \section{Asymptotic $q$-limits of the Coon Amplitude}
\label{A}
In this appendix, we look at the asymptotic $q$-limits of the Coon amplitude.

\subsection*{The $q\to 1$ limit}
To recover the Veneziano amplitude as $q\to 1$, we rewrite the product form of the Coon amplitude (\ref{Coon}) as
\begin{equation}\label{split}
\begin{split}
	\prod_{n=0}^{\infty} \frac{(\sigma\tau-q^n)(1-q^{n+1})}{(\sigma-q^n)(\tau-q^n)}
	&=\frac{(-s-t+2m^2)+(1-q)(-s+m^2)(-t+m^2)}{(-s+m^2)(-t+m^2)(1-q)}\times \prod_{n=1}^{\infty}(1-q^n)\\&\times \prod_{n=1}^{\infty} \frac{(-s-t+2m^2)+(1-q)\,(-s+m^2)(-t+m^2)+[n]_q}{(1-q)[(-s+m^2)+[n]_q][(-t+m^2)+[n]_q]}
\end{split}
\end{equation}
where we have used equation (\ref{sigma_tau}). Next, we note that $$\prod_{n=1}^{\infty}\frac{1-q^n}{1-q}=\prod_{n=1}^{\infty}[n]_q,$$ 
using which we take the limit $q \to 1$ such that \eqref{split} takes the following form:
\begin{equation}\label{one}
	\frac{1}{1-q}\,\frac{x+y}{xy}\prod_{n=1}^{\infty} \frac{n(x+y+n)}{(x+n)(y+n)}=\frac{1}{1-q}\,\frac{\Gamma(x)\Gamma(y)}{\Gamma(x+y)}.
\end{equation}
Here $x=(-s+m^2)$ and $y=(-t+m^2)$, and we have used an identity of beta function in the second equality above. The other factors in (\ref{Coon}) have the limit
\begin{equation}
	(q-1)\,q^{\frac{\log\sigma\,\log\tau}{(\log q)^2}}\to (q-1)
\end{equation}
Combining this with (\ref{one}), we get the familiar Veneziano amplitude in (\ref{limits}) as $q\to 1$. 

\subsection*{The $q\to 0$ limit}

In the limit $q\to 0$, only the $n=0$ factor contributes to equation (\ref{split}), which takes the form
\begin{equation}
\begin{split}
	\prod_{n=0}^{\infty} \frac{(\sigma\tau-q^n)(1-q^{n+1})}{(\sigma-q^n)(\tau-q^n)} &\to \frac{(-s-t+2m^2)+(-s+m^2)(-t+m^2)}{(-s+m^2)(-t+m^2)}\\
	&=\frac{1}{s-m^2}+\frac{1}{t-m^2}-1
	\end{split}
\end{equation}
In the same limit, the pre-factor becomes minus one, and we thus recover the familiar point particle (scalar) result in (\ref{limits}) as $q\to 0$.

\bibliographystyle{JHEP}
\bibliography{citation.bib}
\end{document}